# Structural Discrimination of Phosphate Contact Ion Pairs in Water by Femtosecond 2D-IR Spectroscopy

**Achintya Kundu, Jakob Schauss, Benjamin P. Fingerhut, and Thomas Elsaesser**
*Max-Born-Institut für Nichtlineare Optik und Kurzzeitspektroskopie, Berlin 12489, Germany.*
*Author e-mail address: kundu@mbi-berlin.de*

**Abstract:** The distinct structures of contact ion pairs in water are identified. Nonlinear infrared (IR) spectroscopy and theoretical calculations allow for the separation and assignment of spectral features and interactions. © 2020 The Author(s).

## 1. Introduction

Contact ion pairs and solvent separated ion pairs are fundamental geometries in which ions are accommodated in an aqueous environment [1]. Contact ion pairs are particularly relevant in, e.g., determining the electric conductivity of liquid electrolytes and ionic aqueous solutions and in the stabilization of the biological structures of DNA and RNA [2]. In the latter, the strong Coulomb repulsion between negatively charged phosphate groups of the sugar-phosphate backbone must be overcome via a fine balance of hydration and ion coordination. Nevertheless, the equilibrium properties of aqueous contact ion pairs are challenging to characterize through experiment and simulations, and structural differences among aqueous contact ion pairs are hardly characterized.

Combining linear infrared absorption, femtosecond infrared pump-probe and 2D-IR spectroscopy with atomistic molecular dynamics simulations, we reveal distinct structural features of contact ion pairs consisting of the phosphate group of dimethyl phosphate anion (DMP) and $Na^+$, $Ca^{2+}$, and $Mg^{2+}$ ions. The asymmetric phosphate stretching vibration $\nu_{AS}(PO_2)^-$ serves as most sensitive spectroscopic probe that allows to reveal subtle differences in structure of the different contact pairs. Compared to linear infrared absorption spectra, the nonlinear spectra provide increased sensitivity towards structural differences among contact ion pairs, because the spectra accentuate differences in peak frequency, transition dipole moment strength, and excited state lifetime. The experiments together with long-time molecular dynamics simulations reveal that contact pairs with $Mg^{2+}$ display the most rigid and ordered structure, while pairs with $Ca^{2+}$ and $Na^+$ exhibit larger structural variation and flexibility.

## 2. Results and Discussion

Fig. 1a shows linear infrared absorption spectra of $\nu_{AS}(PO_2)^-$ upon addition of different ions, $Na^+$, $Ca^{2+}$, and $Mg^{2+}$ ions, respectively. The lineshape of the $\nu_{AS}(PO_2)^-$ absorption band of the DMP reference sample exhibits a plateau-like shape that arises from contributions of two conformers of DMP [3]. Upon addition of $Na^+$ ions a slight decrease in intensity of the $\nu_{AS}(PO_2)^-$ absorption band is observed, while for $Ca^{2+}$ and $Mg^{2+}$ ions a pronounced high-frequency shoulder at 1240 and 1250 cm$^{-1}$ can be identified. The blue-shifted shoulder in linear absorption spectra of $\nu_{AS}(PO_2)^-$ has been assigned to contact ion pairs of the $(PO_2)^-$ unit of DMP with doubly charged ions [4,5].

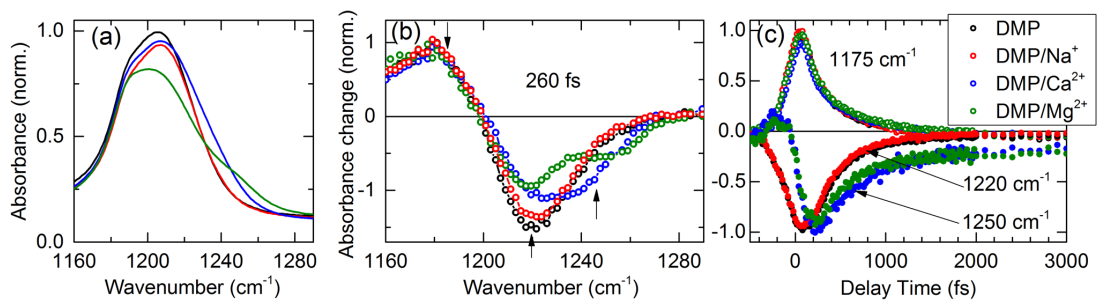

**Figure 1.** (a) Linear absorption spectra of the asymmetric phosphate stretching vibration $\nu_{AS}(PO_2)^-$ of DMP (black line, 0.2 M) and after addition of the 2M excess ions of $Na^+$, $Ca^{2+}$ and $Mg^{2+}$ (red, blue and green lines). (b) Normalized pump-probe spectra at a delay time of 260 fs. The arrows indicate the spectral positions where the kinetic traces are taken. (c) Time resolved absorbance changes at probe frequencies of 1175 cm$^{-1}$, 1220 cm$^{-1}$, and 1250 cm$^{-1}$ as a function of delay time for the four samples.

Fig. 1b presents the normalized pump-probe spectra of the asymmetric phosphate stretching vibration $\nu_{AS}(PO_2)^-$ of DMP (c = 0.2 M) with addition of 2M excess ion concentration at a delay time of 260 fs. The negative signal at 1220 cm$^{-1}$ is due to ground state bleaching and stimulated emission of the v=0→1 vibrational transition and the positive signal at 1175 cm$^{-1}$ arises due to excited state absorption (v=1→2). The pump-probe spectra of the DMP reference sample and DMP with Na$^+$ ions are almost indistinguishable while the pump-probe spectra in presence of Ca$^{2+}$ or Mg$^{2+}$ ions are broadened towards higher frequencies. The reduced signal amplitude of the v=0→1 bleaching and stimulated emission signal arises from spectrally overlapping contributions of DMP species with and without Ca$^{2+}$ or Mg$^{2+}$ ions. For DMP with Mg$^{2+}$ a distinct contribution centered around ~ 1250 cm$^{-1}$ can be identified.

Fig. 1c shows kinetic traces at probe frequency 1175 cm$^{-1}$ (the excited state absorption), 1220 cm$^{-1}$ (maximum of bleaching signal) and 1250 cm$^{-1}$, i.e., at the spectral position of the fundamental transition of contact ion pairs with Ca$^{2+}$ and Mg$^{2+}$ ions. The vibrational lifetime of the asymmetric phosphate stretching vibration $\nu_{AS}(PO_2)^-$ is 370 ± 50 fs (single exponential fit). While the decay at 1220 cm$^{-1}$, i.e., the frequency position of DMP solvated by water is unaffected in the different measurements, the decay at 1250 cm$^{-1}$ is slightly slower at the frequency position assigned to the DMP contact pairs with decay times between 430 and 580 fs.

Fig. 2 displays absorptive 2D-IR spectra of 0.2 M DMP in (a) H$_2$O, (b) 0.2 M DMP in H$_2$O with 2 M Na$^+$, (c) Ca$^{2+}$, and (d) Mg$^{2+}$ ion excess concentration, recorded at a waiting time T=500 fs. The 2D spectrum of DMP in neat H$_2$O (Fig. 2a) displays a single peak on the v=0→1 transition. The elliptic lineshape appears tilted with respect to the excitation frequency axis, reflecting moderate inhomogeneous broadening. Upon addition of Na$^+$ excess ions, the v=0→1 feature of the 2D spectrum broadens along the diagonal which is quantified from diagonal cuts (Fig. 2e,f). For Ca$^{2+}$ and Mg$^{2+}$ excess ions (Figs. 2c,d,g,h) broadening along the diagonal is more pronounced leading to an isolated blue-shifted component in the 2D spectrum. For all investigated ions, 2D cross peaks between the blue-shifted signal contributions and the original DMP band are absent. This finding demonstrates that the underlying vibrations arise from distinct chemical species. Moreover, chemical exchange of species is absent, as would be reflected in the dynamics of cross peaks for varying waiting time T. This indicates that contact ion pairs are structurally preserved on the observation time scale of the experiment. The moderate longer lifetime of $\nu_{AS}(PO_2)^-$ of the contact ion pairs (Fig. 1c) leads to a relative enhancement of the blue-shifted component of the 2D signal allowing for enhanced spectral separation (cf. cuts along the frequency diagonal in Figs. 2g,h).

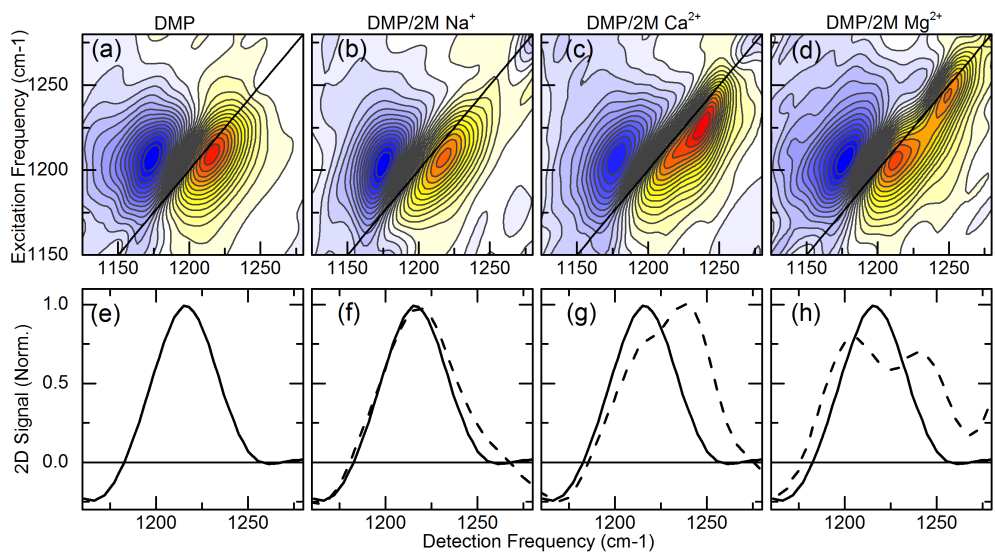

**Figure 2.** 2D-IR spectra of the asymmetric phosphate vibration $\nu_{AS}(PO_2)^-$ measured at a waiting time T=500 fs. (a) DMP in H$_2$O, (b) DMP in H$_2$O with 2 M Na$^+$, (c) DMP in H$_2$O with 2 M Ca$^{2+}$, and (d) DMP in H$_2$O with 2 M Mg$^{2+}$. Absorptive 2D signals are plotted as a function of excitation frequency $\nu_1$ and detection frequency $\nu_3$. Contributions arising from ground state bleaching and stimulated emission of the v = 0 →1 transition are shown with yellow-red contours and v = 1 → 2 excited state absorption contributions are shown in blue. Signal amplitudes are normalized to the maximum ESA signal for the individual 2D spectra. (e-h) Frequency cuts of 2D-IR spectra along a diagonal crossing the maximum of the respective v=0-1 peak (dashed lines), solid lines in (f-h) represents the diagonal cut of the 2D-IR spectrum of DMP in water for reference.

Fig. 3 present results of long-time MD trajectories that analyze contact ion pair geometries of DMP with Na$^+$, Ca$^{2+}$ and Mg$^{2+}$ in a water surrounding. Contact ion pair geometries are analyzed via the two-dimensional potential of mean force (2D-PMF) for the different ions in reduced dimensional coordinate space of the P…ion distance and the P…O1…ion angle α = ∢(Ion$^{x+}$…O1…P) (cf. Fig. 3b,c). The angle α takes a value of 180° for linear arrangements of the P=O group and the ion while α ≈ 90 ° if the ion is placed in the center of the bisector of the (PO$_2$)$^-$ group.

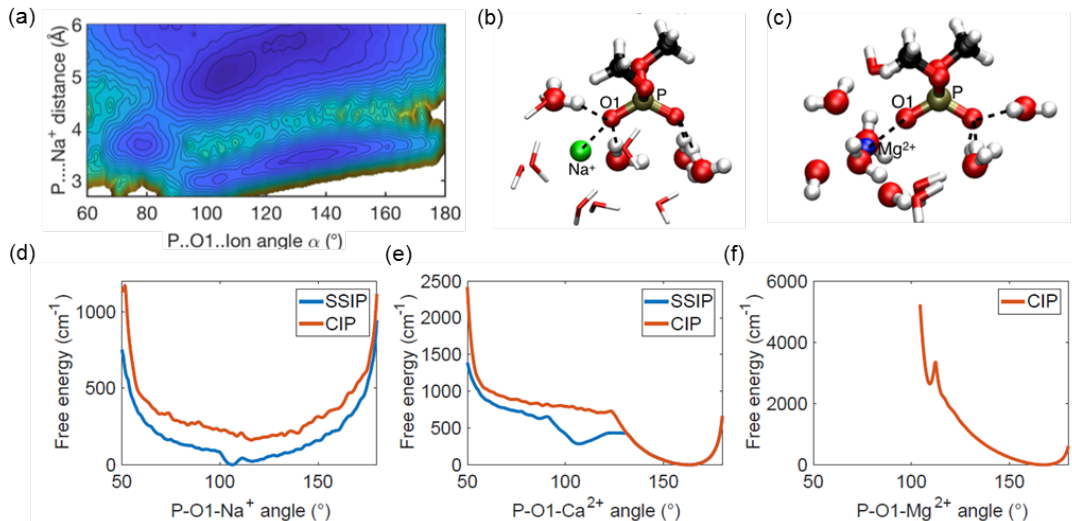

**Figure 3.** Two dimensional potential of mean force (2D-PMF) in reduced dimensional space of the P...Na$^+$ distance and angular coordinate α = ∢(Na$^+$...O1...P) obtained from a 1.59 μs molecular dynamics trajectories (12-6-4 Li-Merz ion parameters and TIP4P-FB water model). (b) Prototypical solvation geometry of a DMP(H$_2$O)$_{11}$Na+ cluster with α~127°. (c) Prototypical solvation geometry of a DMP(H$_2$O)$_{11}$Mg$^{2+}$ cluster for angle α~174°. (d-f) Angular dependence of the free energy of Na$^+$, Ca$^{2+}$ and Mg$^{2+}$ each obtained from a 1.59 μs molecular dynamics trajectory. Minimum energy angular energy profiles were derived from the 2D-PMF upon locating the energy minima of contact ion pairs (CIP) and solvent separated ion pairs (SSIP).

For Na$^+$, solvent separated ion pairs were found to be the most stable species (Fig. 3a,d) while contact ion pairs are characterized (P...Na$^+$ distance ~ 3.0-3.6 Å) and are slightly less stable (~ 200 cm$^{-1}$). For Ca$^{2+}$ (Fig. 3e), the energetic ordering of solvent separated ion pairs and contact ion pairs is reversed and the contact geometry is more stable by about 286 cm$^{-1}$. For Mg$^{2+}$ (Fig. 5f), only contact ion pairs are sampled during the 1.59 μs simulation time. The minimum energy angular profiles of the DMP complexes with Na$^+$, Ca$^{2+}$ and Mg$^{2+}$ reveal that doubly charged Ca$^{2+}$ and Mg$^{2+}$ predominantly form contact ion pairs with an approximate linear arrangement of the P=O group and the ion (α$_{min}$ > 163°, Fig. 3e,f). The O1 atom of the (PO$_2$)$^-$ group occupies the position of one of the water oxygen atoms in the first solvation shell of he doubly charged ion. In contrast, the angular profile of Na$^+$ (Fig. 3d, α$_{min}$ ~ 127°) reflects differences of the contact ion and first solvation shell geometry where Na$^+$ intercalates into the tetrahedral hydrogen bond geometry around the O1 atom of the (PO$_2$)$^-$ group (Fig. 3b).

Our results characterize ion contact pair geometries with unprecedented detail at the molecular level. Characteristic differences for the different ions are identified that reflect different ion charge states (Na$^+$ vs Ca$^{2+}$ and Mg$^{2+}$) and differences in ion radius for a given charge state (Ca$^{2+}$ vs Mg$^{2+}$). Differences in vibrational lifetime of contact ion pairs with Ca$^{2+}$ and Mg$^{2+}$ increase the contrast towards the contact ion pairs in the time resolved measurements making 2D-IR an excellent analytical tool for contact ion pair sensing. Our results demonstrate that the intricate properties of the solvation shell around the phosphate group and the ion are essential to explain the experimental observations.